\documentclass[aps,prx,reprint,amsmath,amssymb,superscriptaddress,nonbibnotes]{revtex4-1}

\usepackage[T1]{fontenc}       
\usepackage[]{graphicx}
\usepackage[utf8]{inputenc}
\usepackage{blindtext}
\usepackage{lmodern}%
\usepackage{hyperref}%
\usepackage{xcolor}%
\usepackage{placeins}
\usepackage{bm}
\usepackage{tabularx}
\usepackage{booktabs}

\newcommand{\cm}{cm\ensuremath{^{-1}}}
\newcommand{\invsqcm}{cm\ensuremath{^{-2}}}
\newcommand{\beq}{\begin{equation}\begin{aligned}}
\newcommand{\eeq}{\end{aligned}\end{equation}}

\newcommand{\bra}[1]{\ensuremath{\langle #1 |\,}}
\newcommand{\ket}[1]{\ensuremath{|#1\rangle\,}}

\newcommand{\mum}{\ensuremath{\mu}m}
\newcommand{\ws}{WS\ensuremath{_2}}
\newcommand{\wse}{WSe\ensuremath{_2}}
\newcommand{\mos}{MoS\ensuremath{_2}}
\newcommand{\tmd}[1]{#1\ensuremath{_2}}
\newcommand{\Ag}{A\ensuremath{_{\rm 1g}}}
\newcommand{\Eg}{E\ensuremath{_{\rm 2g}}}

\newcommand{\degree}{\ensuremath{^\circ\,}}

\newcommand{\boq}{\ensuremath{\mathbf{q}}}
\newcommand{\bok}{\mathbf{k}}

\newcommand{\uFsqcm}{\ensuremath{\rm \mu F\, cm^{-2}}}

\newcommand{\ie}{\emph{i.e.}}

\newcommand{\theos}{Theory and Simulation of Materials (THEOS), and National Centre for Computational Design and Discovery of Novel Materials (MARVEL), \'Ecole Polytechnique F\'ed\'erale de Lausanne, CH-1015 Lausanne, Switzerland}
\newcommand{\dqmp}{Department of Quantum Matter Physics, University of Geneva, 24 Quai Ernest Ansermet, CH-1211 Geneva, Switzerland}
\newcommand{\gap}{Group of Applied Physics, University of Geneva, 24 Quai Ernest Ansermet, CH-1211 Geneva, Switzerland}
\newcommand{\lpmc}{Laboratoire de Physique de la Mati\'ere Complexe (LPMC,  \'Ecole Polytechnique F\'ed\'erale de Lausanne, CH-1015 Lausanne, Switzerland}

\definecolor{linkcol}{rgb}{0,0,0.4}
\definecolor{citecol}{rgb}{0.5,0,0}

\hypersetup{
	bookmarksopen=true,
	pdftitle="Enhanced electron-phonon interaction in multi-valley materials",
	pdfauthor="Ponomarev et al",
	pdfsubject="Enhanced electron-phonon interaction in multi-valley materials", 
	pdfstartview={FitH},    
	pdfmenubar=true, 
	pdfhighlight=/O, 
	colorlinks=true, 
	pdfpagemode=UseNone, 
	pdfpagelayout=SinglePage, 
	pdffitwindow=true, 
	linkcolor=linkcol, 
	citecolor=linkcol, 
	urlcolor=blue
}

\begin{document}
	
	\title{Enhanced electron-phonon interaction in multi-valley materials}

	\author{Evgeniy~Ponomarev}
	\thanks{These authors have equally contributed}
	\affiliation{\dqmp}
	\affiliation{\gap}
	\author{Thibault~Sohier}
	\thanks{These authors have equally contributed}
	\affiliation{\theos}
	\author{Marco~Gibertini}
	\thanks{marco.gibertini@unige.ch}
	\affiliation{\dqmp}
	\affiliation{\theos}
	\author{Helmuth~Berger}
	\affiliation{\lpmc}
	\author{Nicola~Marzari}
	\affiliation{\theos}
	\author{Nicolas~Ubrig}
	\affiliation{\dqmp}
	\affiliation{\gap}
	\author{Alberto~F.~Morpurgo}
	\thanks{alberto.morpurgo@unige.ch}
	\affiliation{\dqmp}
	\affiliation{\gap}
	
	\begin{abstract}
	
We report a new mechanism responsible for the enhancement of electron-phonon coupling in doped semiconductors in which multiple inequivalent valleys are simultaneously populated. We have identified this mechanism through a combined experimental and theoretical investigation of the vibrational properties of atomically thin group-VI semiconducting transition metal dichalcogenides as a function of carrier density. Specifically, we have performed Raman spectroscopy experiments to measure different phonon modes of mono and bilayer \tmd{MoS}, \tmd{WS}, and \tmd{WSe} embedded in ionic-liquid-gated field-effect transistors, and probed vibrations of these systems over a wide range of carrier densities ($\sim5\times10^{13}$ cm$^{-2}$, both for electron and hole accumulation). We find that in-plane modes, such as the \Eg\ mode, are insensitive to the introduction of charge carriers, whereas phonons with a dominant out-of-plane character exhibit strong softening upon electron accumulation, while remaining unaffected upon hole doping. This unexpected --but very pronounced-- electron-hole asymmetry is systematically observed in all mono and bilayers. To understand its origin, we have performed first-principles simulations within a framework that accounts for the reduced dimensionality of the systems and gate-induced charge accumulation. Through this analysis we establish that the phonon softening occurs when multiple inequivalent valleys are populated simultaneously and that the phenomenon is a manifestation of  an aspect of electron-phonon interaction that had not been previously identified. Accordingly, the observed electron-hole asymmetry in the evolution of the Raman spectra originates from the the much larger energy separation between valleys in the valence bands --as compared to the conduction band-- that prevents the population of multiple valleys upon hole accumulation. We discuss the origin of the enhancement of the electron-phonon coupling in terms of a simple physical picture, and conclude that the phenomenon occurs because the population of multiple valleys acts to strongly reduce the efficiency of electrostatic screening for those phonon modes that cause the energy of the inequivalent valleys to oscillate out of phase. The strong enhancement of electron-phonon coupling in the presence of multivalley populations is a robust mechanism that is likely to play an important role in accounting for different physical phenomena and, based on existing experiments, we argue for its relevance in the occurrence of superconductivity --either gate-induced or spontaneous-- in different transition metal dichalcogenides.

	\end{abstract}
	
	\maketitle

	\section{Introduction}

The electronic and elastic properties of solids are determined by the interaction between charge carriers and the elementary excitations associated with the vibration of the crystalline lattice, namely the phonons. In view of its fundamental relevance, the nature of the interaction between electrons and phonons has been investigated in great depth in many different contexts~\cite{Ziman,Grimvall,Mahan}. In the simplest possible terms, electron-phonon interaction can be understood as originating from the electrostatic potential generated by the lattice distortion associated to the ionic displacement in the presence of a phonon. In other words, exciting a phonon in a crystal displaces the charged ions forming the lattice from their equilibrium positions, thereby generating a local charge imbalance and a corresponding electrostatic potential (known as deformation potential~\cite{Bardeen1950,Schockley1950}), which in turns directly acts on all mobile electrons.

Depending on the specific type of material considered and on the level of detail needed for a precise microscopic understanding, the situation can be more complex. For instance, in some systems (among which, graphene)  the coupling between the lattice deformation associated to certain phonon modes and charge carriers is more appropriately described in terms of a vector potential, rather than an electrostatic one~\cite{Pietronero1980,Sohier2014}. In these cases, charge carriers interact with phonons more similarly to the way they interact with a magnetic field~\cite{Vozmediano2010}. Also, the influence of electron-phonon interaction on physical phenomena depends strongly on whether the energy and momentum relaxation of electrons is fast or slow relative to the phonon frequency, corresponding to the so-called adiabatic or anti-adiabatic limit of electron-phonon coupling~\cite{Engelsberg1963,Maksimov1996}.

Irrespective of the specific situation, what is key to understand electron-phonon interaction (and its strength) in solids is electrostatic screening~\cite{Mahan}. Indeed, the deformation potential generated by the excitation of a long wavelength phonon tends to be screened by the spatial redistribution of the electrons present in the system, which self-consistently reduce the strength of the electron-phonon coupling~\cite{Bogulsawski1977,Khan1984,Kartheuser1986}. Screening is expected to be much more effective in the adiabatic regime --since then electrons are able to equilibrate sufficiently fast in response to the ionic motion-- and a much weaker electron-phonon coupling is accordingly expected in this regime as compared to the antiadiabatic one. Conversely, if the nature of the electron-phonon coupling is properly described in terms of a vector potential, as we mentioned it is the case for some of the phonons in graphene~\cite{ferrari_raman_2007,ferrari_raman_2013}, screening cannot influence electron-phonon coupling~\cite{Sohier2014,Pisana2007} (just simply because no spatial distribution of charge carriers can screen a magnetic field). In this case, electron-phonon coupling is expected not to be significantly affected by the presence of charge carriers.

In semiconducting materials hosting only a small density of mobile charges, electrostatic screening is poor and a strong electron-phonon interaction is expected. Addition of charge carriers (e.g., by doping) can drastically improve screening and cause the strength of electron-phonon interaction to decrease, as observed in a variety of semiconductors (e.g., SrTiO$_3$~\cite{kirzhnits_description_1973,van_mechelen_electron-phonon_2008,Wang2016} and other transition metal oxides~\cite{Verdi2017}). For those cases in which phonons are not effectively screened by the presence of mobile charges (see example above), it can also happen that addition of charge carriers leaves the strength of electron-phonon interaction unaffected~\cite{Sohier2014}. In virtually no case, however, the strength of electron-phonon interaction in a semiconductor is expected to increase significantly upon adding charge carriers.

In contrast with this established understanding, here we uncover a yet unidentified mechanism that causes a very significant strengthening --and not a weakening-- of electron-phonon coupling in  atomically-thin semiconductors upon increasing electron density. Our work relies on a joint experimental and theoretical investigation of Raman spectroscopy performed on mono and bilayers of different semiconducting transition metal dichalcogenides (TMDs;  \tmd{MoS}, \tmd{WS}, and \tmd{WSe}) as a function of density of accumulated charge carriers. By integrating all these atomically-thin layers in ionic-liquid-gated field-effect transistors (FETs)  we have measured the evolution of the Raman spectrum as a function of electron and hole density up to approximately 5$\times$10$^{13}$ cm$^{-2}$. In all investigated mono and bilayers, the experiments reveal unambiguously that the electron-phonon interaction systematically softens out-of-plane vibrational modes only when electrons are accumulated in the FET channel, while leaving them unaffected upon hole doping. To identify the origin of this unexpected electron-hole asymmetry we have performed density-functional-theory (DFT) calculations in the most realistic framework currently available, accounting for the reduced dimensionality of the materials and for the presence of a finite density of charge carriers~\cite{Sohier2017}. The calculations allow us to conclude that the observed phonon softening originates from a pronounced increase in the strength of the electron-phonon coupling that occurs whenever the charge carriers in the systems populate simultaneously two inequivalent valleys. Besides accounting for the observed electron-hole asymmetry (as a direct consequence of the different valley structure in the conduction and valence bands of atomically-thin semiconducting TMDs), this finding reveals an aspect of electron-phonon coupling that had not yet been appreciated. Specifically, the strengthening of the electron-phonon coupling is caused by charge transfer between the inequivalent valleys induced by the deformation potential associated to out-of-plane vibrational modes: being local in real space, this inter-valley charge transfer decreases the density of charge that can be displaced, thereby suppressing the effectiveness of  electrostatic screening. The phenomenon --expected to be of rather general validity-- had not been identified so far, most likely because the models commonly used to describe electron-phonon interaction theoretically do not consider the presence of multiple inequivalent valleys.  Equipped with the understanding resulting from the work presented here, we discuss how existing experiments on  superconductivity in TMDs (both gate-induced~\cite{ye_superconducting_2012,jo_electrostatically_2015,lu_full_2018} and spontaneously occurring~\cite{Hamaue1986,Guillamon2008,Xi2016,Ugeda2016,Tsen2016}) provide evidence correlating the enhancement in electron-phonon interaction due to multivalley populations to the occurrence of the superconducting state.

\section{Experimental results}
\label{sec:ExpResults}

\begin{figure}
\centering
\includegraphics[width =1\linewidth]{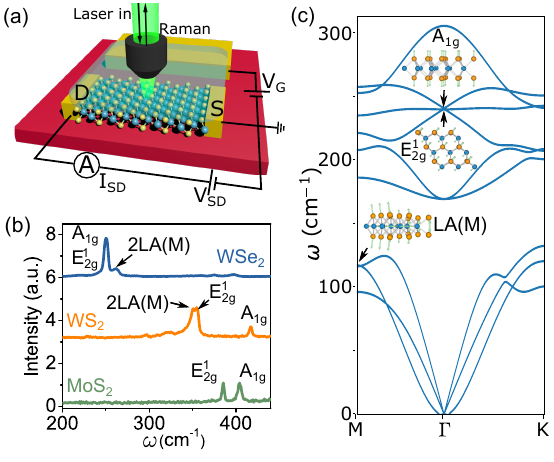}
\caption{(a): Schematics of a ionic-liquid-gated monolayer TMD, showing the biasing configuration employed to operate the device as a field-effect transistor, as well as the microscope objective used to focus the laser light on the device and to couple the light emitted to the spectrometer. (b): Raman spectra measured at room temperature of bare exfoliated monolayers of \tmd{WSe} (blue line), \tmd{WS} (orange line), and \tmd{MoS} (green line). The peaks visible in the curves originate from the \Ag{}, \Eg{}, and 2LA(M) modes, as indicated in the figure. (c): Phonon dispersions of a pristine \tmd{WSe} monolayer in absence of doping obtained by DFT calculations. In \tmd{WSe} the \Ag\ and \Eg\ modes are degenerate at the $\Gamma$-point making them undistinguishable in the Raman spectrum, as can be seen in (b). The diagrams in (c) sketch the atomic displacement pattern for the different Raman active modes identified in (b). }
\label{fig:setup}
\end{figure}

\begin{figure*}
\centering
\includegraphics[width =1\linewidth]{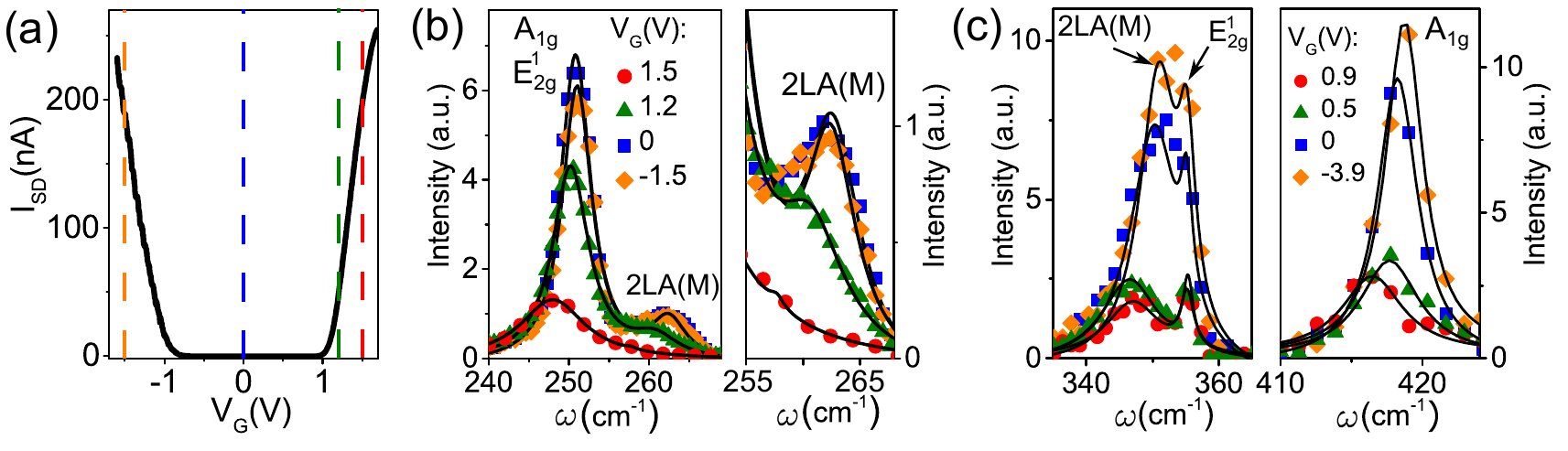}
\caption{(a): Source-drain current I$_{\rm SD}$ as a function of gate voltage V$_{\rm G}$ measured on a \tmd{WSe} monolayer (source-drain bias V$_{\rm SD}$ = 50 mV), exhibiting clear ambipolar transport. This curve is measured \emph{in-situ}, with the device mounted in the set-up used to perform Raman spectroscopy, and it enables us to estimate the density of carriers at any given value of V$_{\rm G}$ as discussed in the main text. (b) Left: Raman spectrum of  \tmd{WSe} monolayer at selected gate voltages indicated in the legend and corresponding to the values of V$_{\rm G}$ marked with the dashed lines of different colours in (a). The peak near 250 cm$^{-1}$ originates from the degenerate \Ag{} and \Eg{} modes, and the one near 250 cm$^{-1}$ is due to a resonant electronic process involving the LA(M) mode. In all cases the black lines are fit of the data to Lorentzian-shaped peaks centered around the energy of the different modes identified. Right: zoom in on the spectral region around the peak due to the resonant process involving the LA(M) mode. (c) Left: Peaks in the Raman spectrum of \tmd{WS} bilayer originating from the \Eg\ and 2LA(M) modes for different values of gate voltage (see legend in the right panel). Right: Peak in the Raman spectrum due to the \Ag\ mode measured at the gate voltages indicated in the legend (black lines are fits of the data obtained by introducing a Lorentzian-shaped peak for each one of the modes involved). As for \tmd{WSe} monolayer the spectra are strongly affected upon electron accumulation; the \Ag\ and 2LA(M) modes shift to lower wavenumbers (i.e. softening) and decrease in intensity, whereas they are left virtually unchanged upon hole accumulation. The position of the \Eg\ peak remains constant for both electron and hole accumulation.}
\label{fig:raman}
\end{figure*}

Our experiments consist in measurements of the Raman spectrum of mono and bilayers of different semiconducting TMDs (\tmd{WSe}, \tmd{WS}, and \tmd{MoS}) as a function of carrier density. The density and polarity of charge carriers is varied continuously by employing these atomically-thin crystals as active parts of ionic liquid gated field-effect transistors (see Fig.~\ref{fig:setup}(a) for a schematic illustration). It has been shown in a multitude of experiments over the last several years that ionic-liquid gating is extremely effective in combination with semiconducting TMDs, as it allows the accumulation of large densities (even in excess of  10$^{14}$ cm$^{-2}$) of both electrons and holes in a same device~\cite{braga_quantitative_2012,ye_superconducting_2012,zhang_electrically_2014,lezama_surface_2014,jo_mono-_2014,ubrig_scanning_2014,ponomarev_ambipolar_2015,jo_electrostatically_2015,zhang_ambipolar_2012,costanzo_gate-induced_2016,gutierrez-lezama_electroluminescence_2016}. One of the characteristic manifestations of the possibility to vary the charge density and type over such a large interval is the occurrence of ambipolar transport, which can be observed upon sweeping the gate voltage V$_{\rm G}$ (see Fig.~\ref{fig:raman}(a) for an example). In the present context, the occurrence of ambipolar transport is useful because it allows us to obtain a fairly accurate estimate of the carrier density in the devices used for the Raman measurements \emph{in-situ}. In practice, in the experiments we record the Raman spectrum at many different values of applied gate voltage and --as we vary V$_{\rm G}$-- we also apply a small bias voltage, V$_{\rm SD}$ between the source and the drain electrode to measure the current, I$_{\rm SD}$, passing through the transistor channel. This allows us to determine the threshold voltage for electron and hole conduction --$V_{th}^{\rm e}$ and $V_{th}^{\rm h}$-- from which we directly estimate the density of electrons and holes at any given value of V$_{\rm G}$ as  $n_e=\frac{C|V_G-V_{th}^{\rm e}|}{e}$ and $n_h=\frac{C|V_{\rm G}-V_{th}^{\rm h}|}{e}$ (where $C$ is the capacitance per unit area obtained  from previous experiments on analogous devices, in which Hall effect measurements were performed to determine the density of charge carriers). We estimate the uncertainty in the value of carrier density extracted in this way to be approximately 20-30~\%, which is sufficient for the purpose of this work (a more precise determination would require measuring the Hall resistance using a superconducting magnet which cannot be done in the same set-up that we use to perform Raman spectroscopy). For more details concerning the fabrication and operation of ionic-liquid FETs based on atomically-thin TMDs we refer the reader to our earlier work (Refs.~\onlinecite{braga_quantitative_2012,jo_mono-_2014,ponomarev_ambipolar_2015}) and to Appendix~\ref{app:methods}.

Characteristic Raman spectra from bare monolayers of WSe$_{2}$ (blue line), WS$_{2}$ (orange line), and MoS$_{2}$ (green line) are shown in Fig.~\ref{fig:setup}(b). We focus on the most prominent features in these spectra, whose  assignment in terms of Raman active vibrational modes has been discussed extensively in the literature~\cite{lee_anomalous_2010,Molina-Sanchez2011,berkdemir_identification_2013,sahin_anomalous_2013,saito_raman_2016}. Using for simplicity the same nomenclature as in the bulk, these are the in-plane (\Eg) and the out-of-plane (\Ag) modes, as indicated in Fig.~\ref{fig:setup}(b), which are seen in all group VI TMDs. For WS$_{2}$ and MoS$_{2}$ these features are spectrally separated while for WSe$_2$ the modes hybridize due to an accidental degeneracy in the phonon dispersion relation, and appear as a single peak at about 250 \cm{}. The degeneracy is also present in the calculated phonon dispersion relation of \tmd{WSe} in Fig.~\ref{fig:setup}(b), on which we point to the modes observed in the Raman spectroscopy data. For most peaks the observed frequency corresponds to the value of the corresponding mode at the $\Gamma$-point. However one of the peaks, visible in the spectra of both \tmd{WSe} and \tmd{WS} and labeled 2LA(M) (see Fig.~\ref{fig:setup}(b)), is due to a double resonant process~\cite{terrones_new_2014,mitioglu_second-order_2014}, which occurs because the frequency of the laser used in our measurements matches an electronic transition  (C absorption for \tmd{WSe} and the B-exciton for \tmd{WS})~\cite{carvalho_band_2013}. Although it is not the main point of interest here, the possibility to observe this mode is relevant because it allows the investigation of the coupling between electrons and phonons at finite momentum, i.e. away from center of the Brillouin zone, as indicated by the corresponding label LA(M) in  Fig.~\ref{fig:setup}(c).

\begin{figure*}
\centering
\includegraphics[width =1\linewidth]{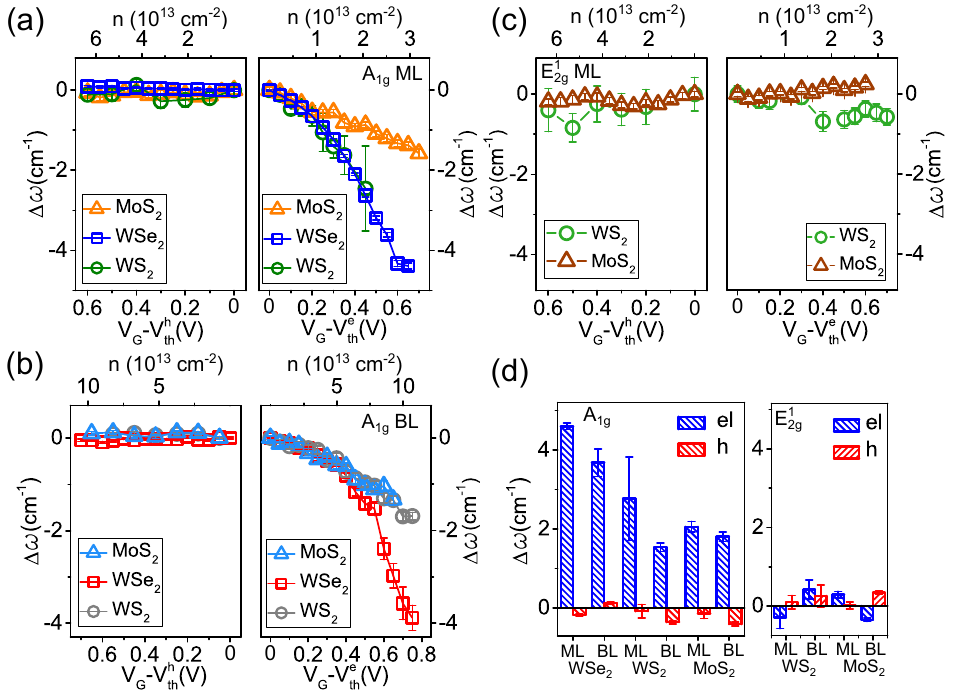}
\caption{(a): Shift $\Delta\omega$ of the \Ag\ peak position in \tmd{MoS} (yellow open triangles), \tmd{WSe} (blue open squares), and \tmd{WS} (green open circles) monolayers as a function of gate voltage (bottom axis), relative to the threshold voltage for hole accumulation (left panel) and for electron accumulation (right panel). The values of electron and hole density corresponding to the applied gate voltage --estimated using the capacitance of the ionic liquid gate, $n_{e,h}= \frac{1}{e}C|V_{\rm G}-V_{\rm th}^{\rm e,h}|$-- is indicated on the top axis (the capacitance values are largely determined by the density of states, as discussed in note~\cite{capa}). The position of the Raman peak originating from the \Ag\ mode remains unchanged for all TMD monolayers upon hole accumulation and shifts (phonon softening) upon electron accumulation. (b): Same as (a) for the \Ag\ mode of bilayers of the same TMD compound, \tmd{MoS} (light blue open triangles), \tmd{WSe} (red open squares), and \tmd{WS} (grey open circles). The trends and the order of magnitude of the frequency softening are identical in mono and bilayers. (c): The \Eg\ peak in monolayer \tmd{MoS} (brown triangle) and \tmd{WS} (green circle) remains constant (i.e., $\Delta\omega \sim 0$ cm$^{-1}$) irrespective of the applied gate voltage (the left panel corresponds to hole accumulation and the right one to electron accumulation; for \tmd{WSe} the investgation of the \Eg\ mode is prevented by the degeneracy with the \Ag\ mode). The same insensitivity of the \Eg\ mode to charge accumulation is seen in bilayers (not shown). (d): To summarize the effect of electron and hole accumulation on the softening of the \Ag\ and \Eg\ phonon modes we compare the absolute shift of the peak position at the highest gate voltage value (relative to the corresponding threshold voltage) that has been reached in measurements for all different mono and bilayer TMDs, $|$V$_{\rm G}$-V$_{\rm th}^{\rm e,h}|$ = 0.6 V. The left and right panels summarize the values measured for the \Ag\ and \Eg\ mode respectively, for either electron (blue bars) or holes (red bars). It is clear from these plots that only the \Ag\ modes (with an out-of-plane motion component of the atomic displacement) are affected by introduction of carriers, and only if electrons are accumulated.}%
\label{fig:exp_softening}
\end{figure*}

To illustrate the results of our Raman spectroscopy measurements as a function of accumulated charge density we start by discussing the case of \tmd{WSe} monolayers. The source-drain current I$_{SD}$ as a function of V$_{\rm G}$ (for V$_{\rm SD}$ = 50 mV) measured in the device used for the Raman measurements, is presented in Fig.~\ref{fig:raman}(a) and exhibits clear ambipolar behavior. The increase of I$_{\rm SD}$ at positive values of V$_{\rm G}$ indicates accumulation of electrons in the FET channel while the increase of I$_{\rm SD}$ at negative V$_{\rm G}$ corresponds to the accumulation of holes. As explained above, we use these measurements to estimate the density of electrons and holes as a function of V$_{\rm G}$. Fig.~\ref{fig:raman}(b) shows the Raman spectra measured for selected values of V$_{\rm G}$ indicated by the vertical dashed lines in Fig.~\ref{fig:raman}(a). For positive values of V$_{\rm G}$ electrons are accumulated (green and red lines, i.e. V$_{\rm G}$ = +1.2 and +1.5 V, corresponding respectively to $n_e \sim 1.5\times10^{13}$  and $2.5\times10^{13}$~\invsqcm{}), for negative V$_{\rm G}$ values the gate accumulates holes (orange line, i.e.  V$_{G}$ = -1.5 V, corresponding to $n_h \sim 6 \times 10^{13}$~\invsqcm{}), whereas for V$_{\rm G}$ = 0 V (blue line) the chemical potential is in the gap and the semiconductor is neutral~\cite{capa}. Upon electron accumulation the Raman peak originating from the hybridized \Ag/\Eg\ mode exhibits a clear softening, shifting towards lower wavenumbers by an amount $\Delta \omega \simeq 3-4$ \cm{}, as well as a decrease in intensity as compared to V$_{\rm G}$ = 0 V case. In contrast, accumulation of holes leaves the Raman spectrum virtually unchanged: the peak does not exhibit any appreciable shift, and only its height decreases slightly. The same trends upon varying V$_{\rm G}$ are observed for the resonant Raman peak originating from the 2LA(M) mode, shown in the right panel of Fig.~\ref{fig:raman}(b)~\cite{2LAm}.

Since in \tmd{WSe} monolayers the degeneracy of the \Eg\ and \Ag\ modes prevents us to study separately the evolution of the two individual Raman peaks upon accumulation of charge carriers, in Fig.~\ref{fig:raman}(c) we show data from bilayer WS$_2$, in which these modes are well separated in frequency. The corresponding Raman peak positions at V$_{\rm G}$ = 0 V are located respectively at $\omega$ = 355 \cm\ --Fig.~\ref{fig:raman}(c), left panel-- and 417 \cm\ --Fig.~\ref{fig:raman}(c), right panel. The same figures also show the Raman spectra measured at selected values of V$_{\rm G}$, in a range of carrier density somewhat larger than for monolayer \tmd{WSe}. For values of V$_{\rm G}$ corresponding to hole accumulation only a negligible shift in position, and at most a small change in amplitude, are seen for all peaks. In contrast, upon electron accumulation, the \Ag\ and the 2LA(M) modes exhibit a clear softening similarly to what is observed in \tmd{WSe} (the position of the \Eg\ mode remains unaffected even for electron accumulation).

Fig.~\ref{fig:raman} reveals a number of clear trends of the Raman active modes in \tmd{WSe} monolayer and \tmd{WS} bilayer, such as the insensitivity of all phonon modes to hole accumulation, and the softening of out-of-plane modes upon electron accumulation. We have performed gate-dependent Raman spectroscopy measurements on mono and bilayers of WSe$_{2}$, \tmd{WS} and \tmd{MoS} following a procedure identical to the one described here above for \tmd{WSe} monolayer and \tmd{WS} bilayer, and found that these trends are generically present in all the investigated atomically thin crystals of semiconducting TMDs. To illustrate and compare quantitatively the results of measurements performed on different systems we plot for each one of them the change $\Delta\omega$ in the frequency of the different phonon modes relative to the value measured at threshold for electron or hole accumulation (\ie{}, for each system we plot $\Delta\omega$ versus $V_{\rm G} - V_{\rm th}^{\rm e}$ and $V_{\rm G} - V_{\rm th}^{\rm h}$). The results of these  measurements  for the \Ag\ modes in all the investigated mono- and bilayers are summarized in Fig.~\ref{fig:exp_softening}(a) and \ref{fig:exp_softening}(b). The data for the \Eg\ modes in monolayers are shown in Fig.~\ref{fig:exp_softening}(c). It is clear from this systematic experimental analysis --which goes well beyond what had been done until now ~\cite{chakraborty_symmetry-dependent_2012,lu_gate-tunable_2017}-- that the same behavior is common to all atomically-thin TMDs.

To quantify the strength of the influence of electron and hole accumulation on the frequency of the different phonons, we plot for all systems investigated the frequency shift of the \Ag{} and \Eg{} modes measured at the highest value of applied gate voltage applied (relative to the corresponding threshold voltage), $|V_{\rm G} - V_{\rm th}^{\rm e,h}| = 0.6$ V, for which we have obtained data for both mono and bilayer, as well as for electron and hole accumulation. Fig.~\ref{fig:exp_softening}(d) demonstrates clearly that, for both mono and bilayers, only Raman active phonon modes with an out-of-plane component of the atomic displacement are influenced by the presence of  charge carriers, whereas the \Eg\ mode --for which the atomic displacement is in-plane-- is left unaffected by the presence of free charge. The data also make the strong electron-hole asymmetry in the softening of the out-of-plane modes very apparent for all the different TMDs investigated. The order of magnitude of the effect --which is what we discuss here-- is the same in all cases, corresponding to a shift of 2-4 \cm{} for the \Ag{} optical Raman mode (for a more detailed quantitative analysis it should be recalled that --since the value of the capacitance used is different for mono and bilayers and for electron and hole accumulation-- the same value of gate voltage relative to threshold does not correspond to the same carrier density~\cite{capa}). 

\section{Theoretical analysis}
\label{sec:theory}

The experiments discussed in the previous section show unambiguously that a strong electron-hole asymmetry in the softening of the phonon frequency is a generic property of atomically thin semiconducting TMDs. The microscopic mechanism at the origin of this asymmetry, however, is not known and, especially for monolayers, finding such a pronounced asymmetry is unexpected. Indeed, the low-energy electronic states in WSe$_2$, WS$_2$, and MoS$_2$ monolayers at the bottom of the conduction band and at the top of the valence bands --centered around the K/K' points-- are often described in terms of massive Dirac fermions, a conceptual framework that may lead one to expect a high degree of symmetry in the physical response upon accumulation of electrons and holes. Even though the very different strength of spin-orbit interaction in the conduction and valence band limits the validity of these considerations, whether and how spin-orbit interaction can cause such a strong asymmetry in the phonon properties upon electron and hole accumulation is very far from obvious.

\begin{figure}
\centering
\includegraphics[width=0.8\linewidth]{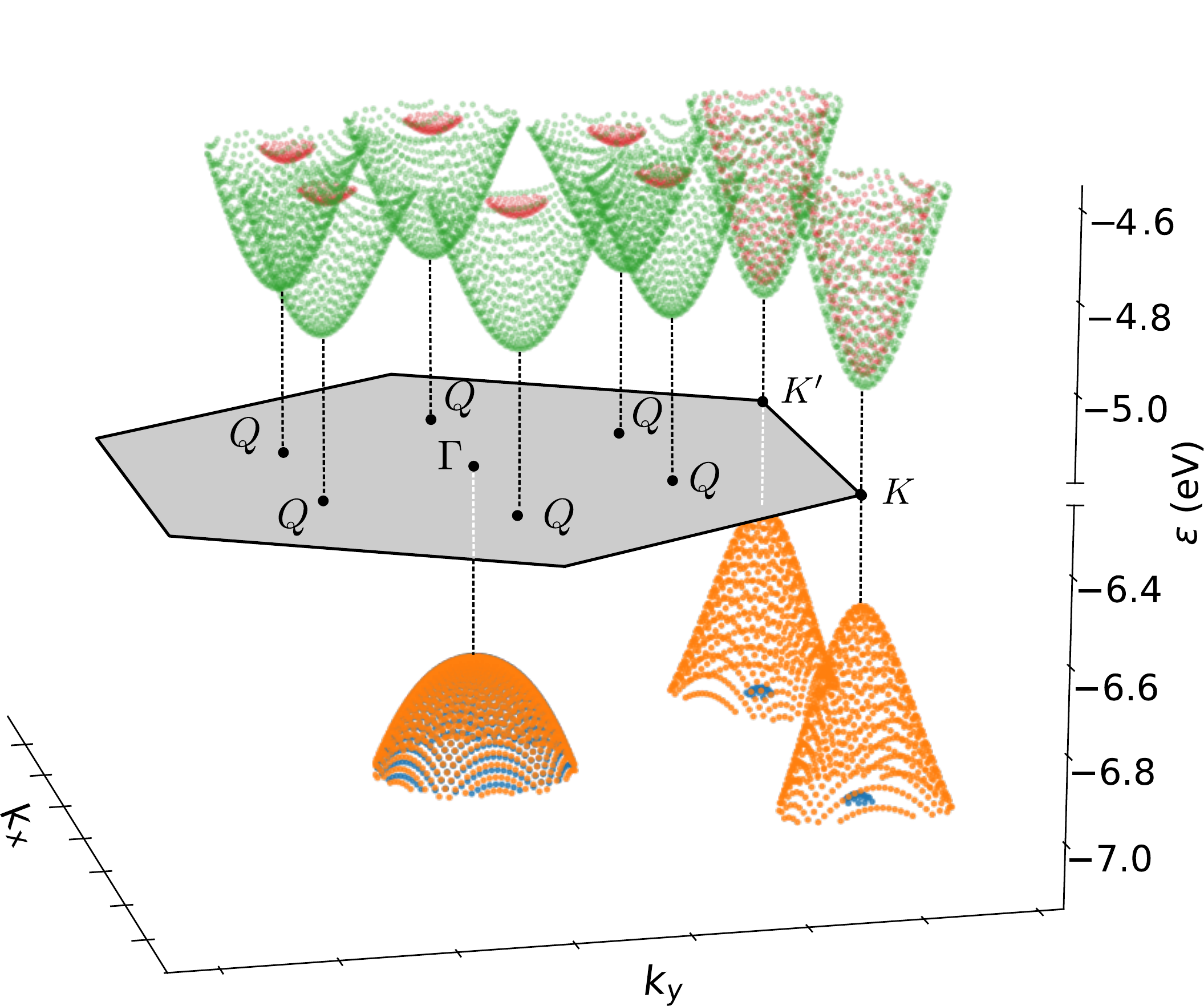}
\caption{The low-energy valley structure of the conduction and valence bands is common to all the TMD monolayers investigated here. This figures illustrates it by showing the low-energy parts of the conduction and valence bands of \ws{} monolayers (the grey shaded hexagon represents the Brillouin zone). The top of the valence band (shown in orange and blue) is formed by a single spin-degenerate valley around $\Gamma$ and a spin-split, two-fold degenerate valley at the K/K$'$ point. The bottom of the conduction bands (represent in green and red) is composed of an almost spin-degenerate valley at the K/K$'$ point (with degeneracy two) and a six-fold degenerate spin-split valley approximately midway between $\Gamma$ and K, normally referred to as the Q point.}
\label{fig:bands}
\end{figure}

The purpose of this section is to present a theoretical analysis showing that the strength of the electron-phonon interaction in atomically-thin semiconducting TMDs is governed by whether or not multiple valleys  (the $\Gamma$ and K valleys in the valence band and the K and Q valleys in the conduction band, see Fig.~\ref{fig:bands}) are simultaneously populated by the accumulated charge carriers. In other words, it is the rather different valley structure of the conduction and the valence bands that is ultimately responsible for the difference in the coupling to phonons of electrons and holes, and that causes the observed electron-hole asymmetry in the softening of the phonon frequencies. In this regard,  spin-orbit interaction does play a prime role, because its strongly asymmetric strength between valence and conduction bands governs the energetic alignment of the different valleys, and therefore determines whether --at any given value of carrier density-- one or multiple valleys are populated.

This conclusion is based on DFT and density-functional perturbation theory (DFPT) calculations, which provide an extremely powerful tool to identify all trends in the observed dependence of the phonon frequency on the density of accumulated charge. DFT, however, cannot reproduce the experiments at a detailed quantitative level. This is because a full quantitative agreement would require a precise prediction of the relative energetic alignment of the valleys present in the valence and conduction band, which is well-known to be extremely sensitive to multiple parameters~\cite{MolinaSanchez2015,Le2015,Kormanyos2015,Yuan2016}, from the choice of exchange-correlation functional to details of the  crystal structure. Indeed, even experimental attempts to determine precisely the energetics of the different valleys seem to give scattered results, depending --for instance-- on the substrate employed in the experiments (see Table~\ref{tab:th_exp}). The situation is even more complex for bilayers, where the electric field associated with the FET doping influences the relative energy of the valleys in a way that cannot be  precisely predicted at this stage. As such, our analysis aims exclusively at exploiting the possibility to tune parameters and conditions in the controlled environment of DFT simulations  to identify the physical processes that explain our experimental results, and not at reproducing in quantitative detail the observed carrier density dependence of the phonon frequencies. That is also why --in view of the subtle effects caused by the gate electric field  in bilayers-- we focus our theoretical analysis exclusively on TMD monolayers.

To start discussing the effect of the accumulated carrier density on the phonons, we note that the softening observed in the experiments is expected when describing electron-phonon interactions within the Born-Oppenheimer approximation. In such an approximation scheme, electrons are assumed to follow the ionic motion, remaining at all times in the ground state corresponding to the instantaneous lattice configuration. This implies that phonons can be considered as static perturbations acting on the electrons, which in turn affect the vibrational properties of the system. In this regime, the expression for the change in  frequency of a phonon at $\Gamma$ with respect to the neutral semiconducting case reads (see appendix~\ref{app:softening} and Refs. \cite{Lazzeri2006,Saitta2008,Calandra2010}):
\begin{align} \label{eq:Dom}
\Delta\omega_\nu \approx  - N(\varepsilon_F)\langle g_\nu^2 \rangle_{FS}.
\end{align}
Here we assume the limit of zero temperature, $N(\varepsilon_F)$ is the density of states at the Fermi energy, $\varepsilon_F$, and $\langle g_\nu^2 \rangle_{FS}$ is the screened electron-phonon coupling (squared) for the $\nu$-th phonon mode averaged over Fermi surface (see Eq.~\eqref{eq:g2aver} in appendix~\ref{app:softening}). It follows that softening ($\Delta\omega_\nu<0$) is expected for phonon modes that are strongly coupled with the electronic states on the Fermi surface, whereas no frequency shift is expected for modes with a negligible electron-phonon coupling. For instance, this is the case of the pure \Eg\ longitudinal optical mode, which affects electrons by generating a long-range scalar potential (known as Fr\"ohlich interaction) that in the presence of free carriers (either electrons or holes) is perfectly screened in the long-wavelength limit, so that $\langle g_\nu^2 \rangle_{FS}\approx0$.

The assumption that carriers follow the ionic motion remaining in the instantaneous ground state implies that the Born-Oppenheimer approximation is a good description only for systems that are in the adiabatic limit. This requires the phonon energy $\hbar\omega_\nu$  to be much smaller than the carrier relaxation rate $\hbar/\tau$ ($\omega_\nu\tau\ll 1$), where $\tau$ is the carrier momentum-relaxation lifetime. In ionic-liquid gated FETs based on monolayer semiconducting TMDs the value of $\tau$ can be  estimated from the experimentally determined carrier mobility that --under the conditions of the experiments (i.e., with ionic-liquid gated devices)-- is typically $\mu \approx$ 15 cm$^2$/Vs for both electrons and holes. When compared to the characteristic phonon frequencies probed in our experiments, the resulting value of $\tau \approx$ 10 fs gives $\omega_\nu\tau \leq 1$. The condition to treat electron-phonon interaction in monolayer TMDs using the Born-Oppenheimer approximation is therefore fairly well satisfied at room temperature. Note also that, even when $\omega_\nu\tau \simeq 1$, theory allows the phonon softening to be estimated as~\cite{Maksimov1996,Marsiglio1992,Saitta2008}:
\begin{equation}\label{eq:Dom_tau}
\Delta \omega^{(\tau)}_\nu \approx \frac{\Delta \omega_\nu}{1+(\omega_\nu\tau)^2}
\end{equation}
where $\Delta\omega_\nu$ is given by Eq.~\eqref{eq:Dom}. It follows directly from this expression that in the regime of our experiments the shift of the phonon frequency upon accumulation of charge carriers is correctly captured by the value in the fully adiabatic limit within a factor of two.

\begin{table}
\caption{Energy separation $\Delta E_{\rm \Gamma K}$ between the top of the valence band at K and $\Gamma$  in the monolayer TMDs investigated in this work. Experimental results are taken from angle-resolved photo-emission spectroscopy (ARPES) measurements in the literature, while DFT data refer to calculations done in the present work. 
}
{\setcitestyle{super}
\renewcommand{\arraystretch}{1.3}
\begin{tabularx}{\linewidth}{l >{\centering\arraybackslash}X >{\centering\arraybackslash}X >{\centering\arraybackslash}X }
\toprule
  & \multicolumn{3}{c}{$\Delta E_{\rm \Gamma K}$ (meV)}\\
  \cmidrule{2-4}
  & \mos\  & \ws\ & \wse\ \\
 \midrule
Exp. & 150\cite{Jin2013}, 310 \cite{Miwa2015,Yuan2016} & 510\cite{Dendzik2015}, 240\cite{Kastl2018},300\cite{Henck2018} & 560\cite{Zhang2016}, 500\cite{Wilson2017}, 890\cite{Le2015}  \\
DFT (this work) & 60 & 230 & 490 \\
\bottomrule
\end{tabularx}
}
\label{tab:th_exp}
\end{table}

The validity of the adiabatic Born-Oppenheimer approximation makes DFT ideally suited to calculate the phonon frequency as a function of carrier density, because in DFT calculations the ground-state density is obtained by keeping the ionic positions fixed. We thus first investigate which phonon modes are  sensitive to accumulation of charge carriers by computing phonon dispersions of neutral, electron-doped and hole-doped monolayer TMDs using DFPT~\cite{Baroni2001}. Since simulations require overall charge neutrality, the accumulated charge is compensated by the presence of gate electrodes, and the correct boundary conditions for 2D materials are provided by cutting off the Coulomb interactions in the direction perpendicular to the layers~\cite{Sohier2017}. In these initial calculations, spin-orbit coupling is not included and a relatively large electronic temperature is used to smear the Fermi surface (see methods in appendix~\ref{app:methods}).

\begin{figure}
\centering
\includegraphics[width=\linewidth]{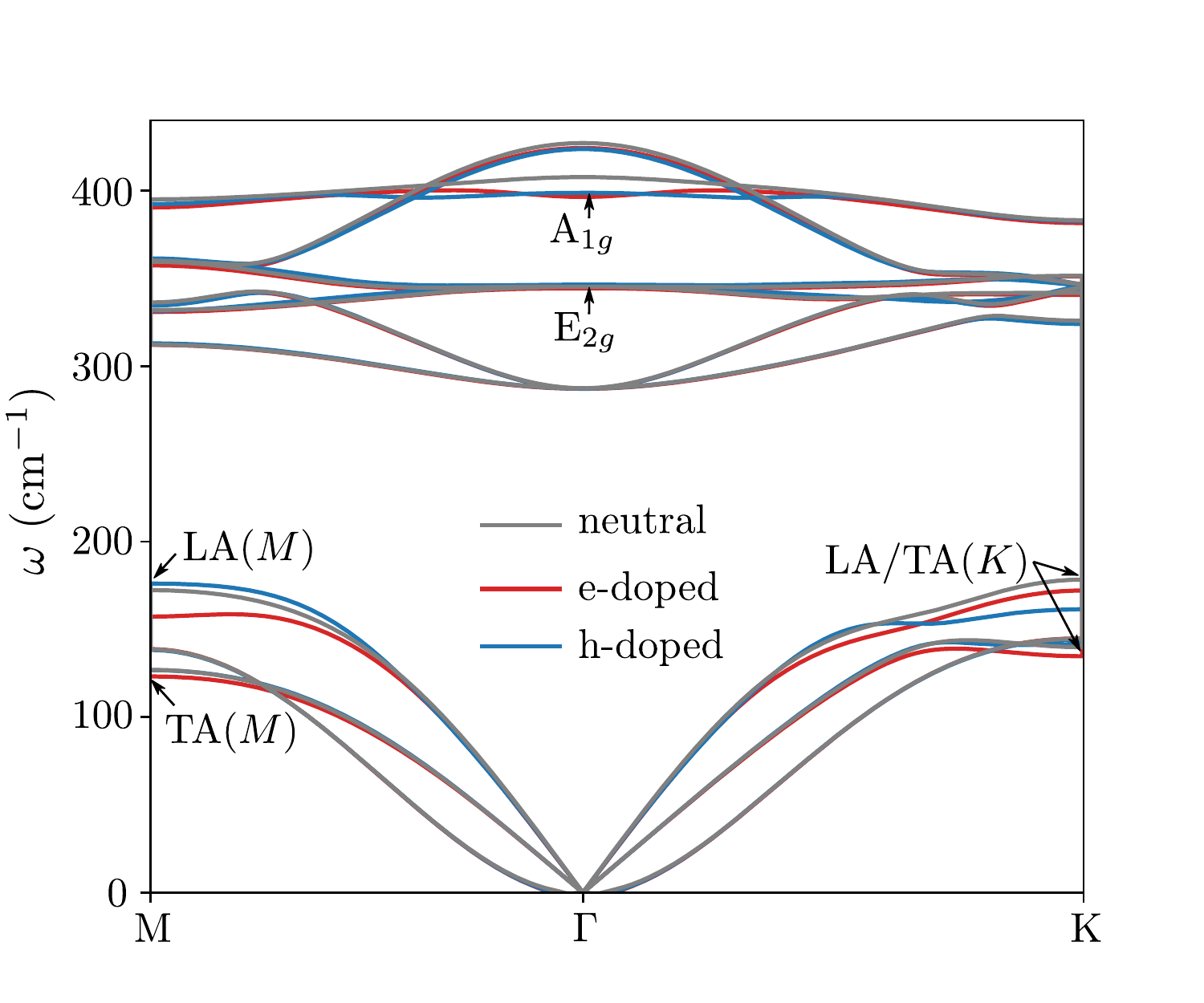}
\caption{Dispersion relations of the different phonons branches, calculated as described in the main text, along the high-symmetry path M-$\Gamma$-K of monolayer \ws. Lines of different colours correspond to either neutral (grey lines) monolayers, or monolayers under electron (blue lines), and hole (red lines) accumulation (calculations are done for 0.02 electrons or holes per unit cell).}
\label{fig:phonon_disp}
\end{figure}

Representative DFPT results are shown in Fig.~\ref{fig:phonon_disp} for \ws. When a finite density of carriers is included, we observe a softening in various branches at several phonon wave vectors. In particular, we see that the \Ag\ mode softens at $\Gamma$ both for electron and hole accumulation, and at M for electron accumulation only. Similarly, the frequencies of the in-plane acoustic modes (LA and TA) decrease at K both upon electron and hole accumulation, and at M exclusively for electron accumulation. On the contrary, the \Eg\ mode remains unaffected. For all the phonon modes detected in the experiments, therefore, the simulations succeed in reproducing which modes are affected by the accumulation of charge carriers, and correctly captures the fact that carrier accumulation causes softening of the phonon frequencies. However, the overall agreement between calculations and experiments is poor. The magnitude of the softening is largely overestimated, well beyond the factor $[1+(\omega_\nu\tau)^2]$ accounting for deviations from the purely adiabatic approximation of DFT. In addition, softening appears also upon hole-doping for the A$_{1g}$ mode at $\Gamma$, in contrast with the experimental results.

\begin{figure*}
\centering
\includegraphics[width=\textwidth]{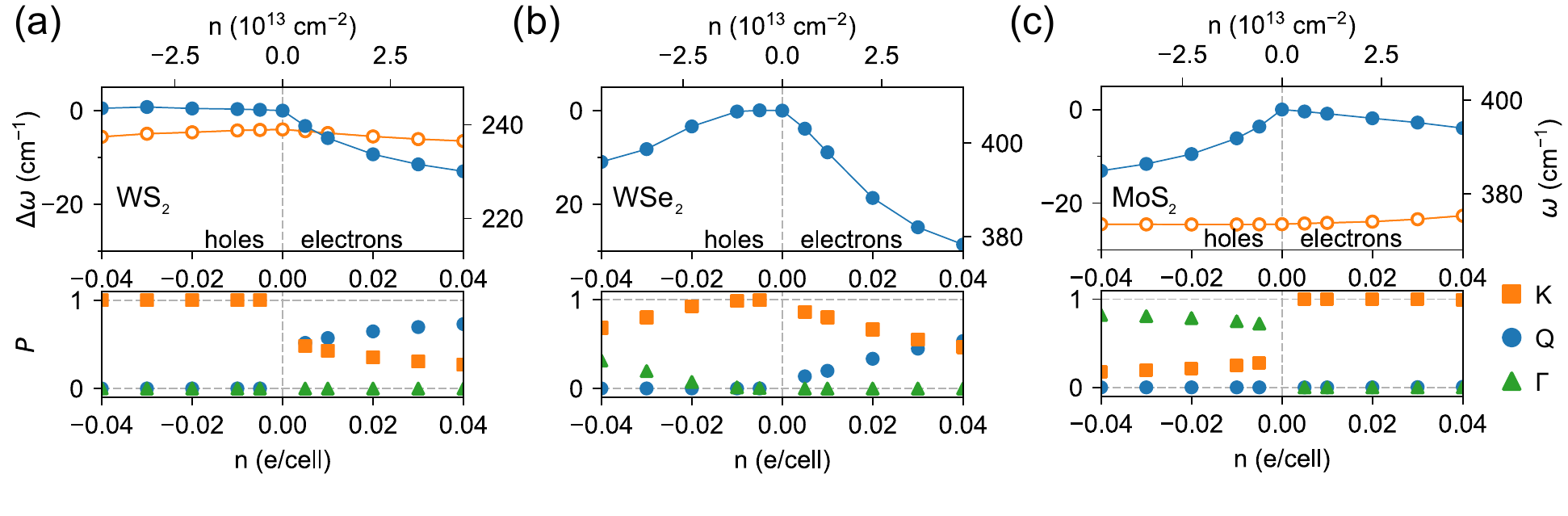}
\caption{ Top panels: Results of ab-initio calculations showing the dependence of the \Ag\ mode frequency (blue circles) on carrier density $n$ in monolayer \ws\ (a), \wse\ (b), and \mos\ (c), for either electron ($n>0$) or hole ($n<0$) accumulation, as indicated in each panel. The range of carrier density considered in the calculations roughly corresponds to that explored in the experiments (see Fig.~\ref{fig:raman}(a)). For \wse\ and \mos\ we also plot the results for the \Eg\ mode (orange empty circles), which is found not to shift, in agreement with experiments (for \ws\ the \Eg\ mode is degenerate with the \Ag\ mode which experimentally prevents its separate determination upon varying carrier density). Lower panels: relative fractional occupation $P$ of the different valleys upon hole and electron accumulation, as extracted from ab-initio calculations done to obtain the density dependence of the phonon frequencies shown in the corresponding top panels. Different symbols refer to the K (orange squares), Q (blue circles), and $\Gamma$ (green triangles) valleys. Note how in all cases --i.e., for all the different monolayers and upon both electron and hole accumulation-- there is a perfect correlation between the shift of the \Ag\ mode and the multiple occupation of valleys (i.e., the K and $\Gamma$ valley in the valence band upon hole accumulation, and the K and Q valley in the conduction band upon electron accumulation).}
\label{fig:A1g_freqs}
\end{figure*}

To investigate the origin of the discrepancy between measurements and simulations, we focus our attention on the \Ag\, which is characterized by an out-of-plane displacement of the chalcogen atoms as illustrated in Fig.~\ref{fig:setup}(b)~\cite{2LAm}. We computed the \Ag\ frequency at $\Gamma$ as a function of carrier density using a finite-difference DFT scheme (see appendix~\ref{app:methods}) for the three monolayer TMDs investigated here. At this stage, we made calculations more realistic by including spin-orbit coupling through fully-relativistic pseudopotentials, and by describing the occupation of electronic states according to the room-temperature Fermi-Dirac distribution (for which we made a much denser sampling of the Brillouin zone). Again, all calculations of total energy and forces include a cutoff of the Coulomb interaction to implement the correct periodic boundary conditions, and gates to maintain overall charge neutrality and properly simulate a FET configuration.

Results for monolayers of the three different TMDs investigated experimentally are shown in Fig.~\ref{fig:A1g_freqs}, both for electron and hole accumulation, in a range of carrier densities comparable to that explored in the experiments. In contrast to the very systematic results of the Raman measurements (see Fig.~\ref{fig:exp_softening}), the evolution of the calculated phonon frequency upon electron and hole accumulation is different for the individual TMDs, and appears at first sight to offer no systematics. In particular, we find that for \ws\ (Fig.~\ref{fig:A1g_freqs}(a)) a large softening occurs on the electrons side, but the \Ag\ mode frequency decreases also upon hole doping (see also Fig.~\ref{fig:phonon_disp}). For \wse\ (Fig.~\ref{fig:A1g_freqs}(b)) instead a significant softening is present only for electrons and not for holes, whereas the opposite is true for \mos\ (Fig.~\ref{fig:A1g_freqs}(c)). Despite this seemingly erratic behavior, we succeeded to identify common trends enabling a rationalization of the results of the calculations. To this end, it is essential to realize that the nature of the electronic states involved plays an important role, because the softening of the phonon frequencies arises from the coupling of phonons with electrons on the Fermi surface (see Eq.~\eqref{eq:Dom}). That is why in the bottom panels of Fig.~\ref{fig:A1g_freqs} we also plot how charge carriers in the different TMD monolayers are distributed among the relevant valleys at K, Q and $\Gamma$ (see Fig.~\ref{fig:bands}), by showing their fractional occupation $P$.

A comparison of the evolution of the phonon frequency with the population of the different valleys does indeed reveal a systematic behavior common to monolayers of all different TMDs, which holds true for both electron and hole accumulation. In all cases a strong softening of the \Ag\ mode occurs only when two inequivalent valleys are simultaneously occupied, whereas the frequency of this mode remains constant whenever the accumulated  carriers occupy exclusively one of the valleys. Specifically, the rapid decrease of the \Ag\ frequency upon electron doping in \ws\ and \wse\ originates from the simultaneous occupation of the K and Q valleys (see orange squares and blue circles in the bottom panels of Fig.~\ref{fig:A1g_freqs} (a) and (b), respectively). Consistently with this idea, in \mos\ only the K-valley is populated, and no significant softening is observed. Analogous trends can be seen upon hole accumulation: A large softening is present in \mos, for which the K and $\Gamma$ valleys are simultaneously occupied, whereas no softening is visible in \wse\ where holes populate only the K valley (see Fig.~\ref{fig:A1g_freqs}(c) and (b), respectively). The case of \ws\ monolayers (Fig.~\ref{fig:A1g_freqs}(a)) confirms once more the underlying notion. In this case, the frequency of the \Ag\ mode is insensitive to hole doping for small concentrations --when only the K valley is occupied-- and starts to soften past 0.02 holes/unit cell, when holes begin to populate both the K and the $\Gamma$ valley. In all cases a very pronounced softening --larger than 10 cm$^{-1}$ at $|n|=5\times10^{13}$~cm$^{-2}$-- is always seen when two distinct valleys are populated, whereas whenever only one valley is populated the change in phonon frequency is typically at least one order of magnitude smaller.

\begin{figure}
\centering
\includegraphics[width=\linewidth]{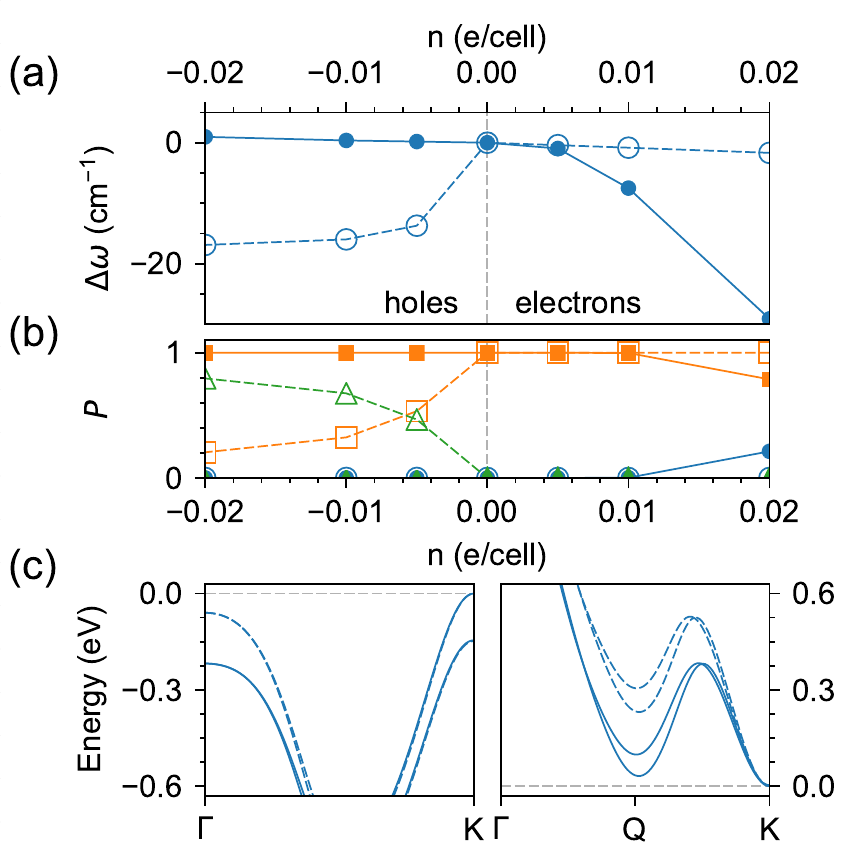}
\caption{Effect of the crystal structure of monolayer \mos\ on  the softening of the \Ag\ mode (a)  and on the fractional occupation $P$ of the K (squares), Q (circles), and $\Gamma$ (triangles) valleys (b) upon carrier accumulation. Empty symbols represent quantities obtained by means of ab-initio calculations performed using the fully relaxed DFT structure of monolayer \mos\, whereas full symbols represent the same quantities obtained using the experimentally-reported structure. Panel (c): band structure of undoped \mos\ close to the top of the valence bands (left) or to the bottom of the conduction bands (right) calculated using the experimentally-reported (continuous lines) or the DFT-relaxed (dashed lines) crystal structure. Note how the different structures very significantly change the distance in energy between the top of the K and $\Gamma$ valleys in the valence band and between the bottom of the K and Q valley in the conduction band. As discussed in the main text, this distance in energy determines whether or not multiple valleys are occupied upon electron or hole accumulation in the calculations, with a corresponding strong effect on the shift of the \Ag\ phonon frequency.
\label{fig:exp_structure}}
\end{figure}

It follows from these considerations that the energy separation between valleys, which governs their relative occupation, plays an essential role in determining the magnitude of the shift in phonon frequencies (i.e., the strength of the electron-phonon coupling) upon carrier accumulation. To confirm this conclusion --and to provide further support for the effect of multivalley populations-- we have performed additional calculations exploiting the extreme sensitivity of TMDs' band structures to the in-plane lattice parameter and the vertical positions of the chalcogen atoms~\cite{Yuan2016}. The goal of these calculations is to artificially tune the energy separation  between valleys in order to check whether, as we expect, the softening of the \Ag\ mode is affected significantly. We have analyzed the behavior of \mos\ monolayers using the experimentally-reported crystal structure (rather than the fully relaxed one obtained by minimizing forces and stresses at the DFT level), for which the energy separation $\Delta E_{\rm \Gamma K}$ between the K and $\Gamma$ valleys in the valence band increases from 60~meV to 220~meV (see Fig.~\ref{fig:exp_structure}(c)), and the energy difference $\Delta E_{\rm K Q}$ between the K and Q valleys in the conduction band decreases from 230~meV to 30~meV. To reduce the spurious effects of thermal broadening in these calculations, we intentionally determine band occupations by using a cold-smeared distribution~\cite{Marzari1999} with $T=150$~K, instead of a broad Fermi-Dirac distribution at $T=300$~K as above.

The outcome of the calculations is illustrated in Fig.~\ref{fig:exp_structure}, where we compare the carrier-density dependence of (a) the frequency of the \Ag\ mode and (b) the fractional valley occupations obtained with either the DFT-relaxed structure (empty symbols, see also Fig.~\ref{fig:A1g_freqs}(c)) or the experimentally-reported one (full symbols). It is apparent that the results are distinctly different in the two cases. Upon hole accumulation a large frequency shift and a simultaneous population of the K and $\Gamma$ valleys are present for calculations performed using the DFT-relaxed structure. In contrast, no shift in frequency occurs in the calculations performed with the experimentally-reported crystal structure, where the fractional occupation of the K valley remains equal to one, i.e. only the K valley is populated. This is exactly the behavior that we would have anticipated as a result of the increased $\Delta E_{\rm \Gamma K}$   for the experimentally-reported crystal structure. Consistently, the reduction in $\Delta E_{\rm K Q}$ (see Fig.~\ref{fig:exp_structure}(c)) in the experimentally-reported structure leads to a non-negligible occupation of the Q valley in addition to the lowest-lying K valley, causing a very large enhancement of the \Ag\ softening upon electron accumulation.

We therefore conclude that the scenario hypothesized above is correct: for all monolayers investigated here, and for both electron and hole accumulation, a very sizable softening of the \Ag\ mode occurs only when charge carriers populate multiple inequivalent valleys. Through Eq.~\eqref{eq:Dom}, this directly implies that also the strength of the electron-phonon coupling for the \Ag\ mode in semiconducting TMDs is large only when two valleys are occupied. This conclusion is particularly interesting because such a strong dependence of the electron-phonon coupling strength on the population of multiple inequivalent valleys in an individual band had not been identified earlier. It is nevertheless extremely likely that this mechanism is at play in many other material systems besides semiconducting TMDs, and relevant to explain a variety of physical phenomena (see, e.g., the discussion of superconductivity in section~\ref{sec:discussion}).

The insight gained on the importance of the simultaneous occupation of multiple valleys can now be exploited to discuss in more detail the results of our Raman spectroscopy experiments (see Fig.~\ref{fig:exp_softening}). To start with, the complete absence of  phonon softening upon hole accumulation indicates that $\Delta E_{\rm \Gamma K}$ is sufficiently large in TMD monolayers to avoid the partial occupation of the $\Gamma$ valley (in addition to the dominant K valley), for hole concentrations as large as $ \sim 5 \times 10^{13}$~cm$^{-2}$. According to the same logic, the phonon softening observed upon electron accumulation indicates that $\Delta E_{\rm KQ}$ in the conduction band is sufficiently small, so that --at room temperature-- both the K and Q valleys are populated in monolayers of all TMDs upon accumulation of $\sim 10^{-2}$ electrons per unit cell. Additionally, it is clear from Fig.~\ref{fig:exp_softening}(a) that phonon softening upon electron accumulation is much more pronounced in W-based TMD monolayers, indicating that $\Delta E_{\rm KQ}$ is smaller for \ws\ and \wse\ than for \mos.

All these conclusions are fully consistent with what is known about monolayers of semiconducting TMDs. The energy difference $\Delta E_{\rm \Gamma K}$ reported in Tab.~\ref{tab:th_exp} from ARPES measurements is consistent with having only the K valley occupied upon accumulation of a density of holes reaching up to $\sim 5 \times 10^{13}$ cm$^{-2}$ ~\cite{Zhang2016,Gao2017,Katoch2018}. Similarly, although no sufficiently systematic ARPES study of the conduction band of monolayer TMDs has been reported yet, analogous measurements on the doped surface of bulk TMDs (where doping should be limited to the first few layers) have shown~\cite{Kang2017} that it is relatively easy to populate both the K and Q valleys in the conduction band (i.e., $\Delta E_{\rm KQ}$ in monolayer TMDs is significantly smaller than $\Delta E_{\rm\Gamma K}$). Moreover, the observation of a larger softening in W-based monolayers, associated with the K and Q valleys being closer in \ws\ and \wse\ than in \mos , is compatible with the fact that $\Delta E_{\rm KQ}$ is largely controlled by spin-orbit coupling (stronger in W-based compounds).

These considerations also make clear why the strong sensitivity of the phonon frequencies to the precise energy separation between valleys severely affects the ability of DFT to make reliable quantitative predictions. In particular, the spurious softening predicted on the hole side of \mos\ (see Fig.~\ref{fig:A1g_freqs}(c)) originates from the large underestimation of $\Delta E_{\rm \Gamma K}$ in DFT as compared to experiments (see Tab.~\ref{tab:th_exp}), which leads to an erroneous occupation of both the K and $\Gamma$ valleys even at very small doping concentrations at room temperature. Something similar, although not as severe, happens also for \ws, for which the DFT prediction for $\Delta E_{\rm \Gamma K}$ is not much smaller than the experimental results in Tab.~\ref{tab:th_exp}, but it is still sufficient to lead to a spurious softening at large hole doping. Finally, the discrepancy with experiments might arise also from a partial failure of DFT  in accounting for band-structure renormalization effects that influence the doping dependence of $\Delta E_{\rm \Gamma K}$ and $\Delta E_{\rm KQ}$.

\section{Discussion}
\label{sec:discussion}
\begin{figure}
\centering
\includegraphics[width=\linewidth]{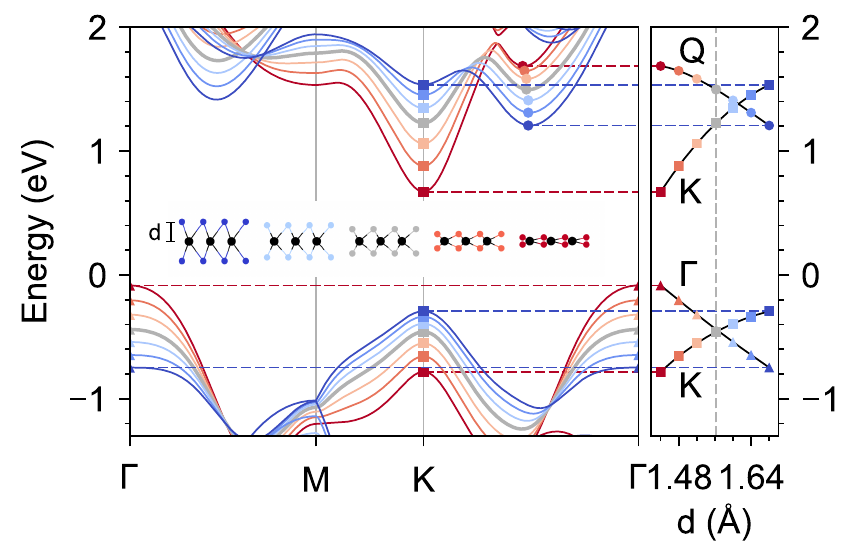}
\caption{Variation of the first-principles energy bands of monolayer \mos\ induced by the displacement pattern of the \Ag\ mode at $\Gamma$ (a rigid vertical motion of the chalcogen planes in opposite directions, with a variation of the distance $d$ between the plane of sulphur atoms and that of molybdenum atoms  -- see inset). To emphasize the effect, the amplitude of the displacement has been set to 0.12~\AA, both in compression (red) and dilation (blue), with respect to the equilibrium value $d = 1.57$~\AA{} (gray). On the right we report the evolution of the band edges of the K and Q valley in conduction and of the $\Gamma$ and K valley in valence, as a function of $d$. The same colours as in the left panel have been used to identify different values of $d$. The absolute value of the slope of each curve at the equilibrium value of $d$ (marked with a vertical dashed line) is proportional to the bare electron phonon coupling of  the corresponding band with the \Ag\ mode at $\Gamma$.
\label{fig:band_ph}}
\end{figure}

Finding that the strength of the coupling of electrons to phonons is drastically amplified when multiple inequivalent valleys  in a given band are populated is a phenomenon that had not been identified in the past, and it is important to build a robust physical understanding. To this end, recall that electron-phonon coupling can be qualitatively understood by considering how the band structure is perturbed by the displacement pattern of the ions associated to the phonon under consideration~\cite{Bardeen1950,Schockley1950,Khan1984}. Fig.~\ref{fig:band_ph} shows the effect of the \Ag\ mode on the band structure of undoped monolayer \mos\ (for better clarity, spin-orbit coupling is not included; we checked that this has no qualitative consequences on the arguments here below). Although the amplitude (up to 0.12~\AA) of the \Ag\ mode has been intentionally exaggerated with respect to typical room-temperature mean-square displacements to illustrate this effect, the large variation of both the valence and conduction bands signals the presence of a very strong electron-phonon coupling. This strength can be quantified by taking the first derivative of the band energy reported with respect to the atomic displacement, which is proportional to the bare electron phonon coupling for that band (see the right panel of Fig.~\ref{fig:band_ph}).

The introduction of free carriers --in our experiments by means of field-effect doping-- allows for the possibility of electrostatic screening, and generically tends to decrease the strength of the electron-phonon coupling (as discussed in Section I). In a simple Thomas-Fermi picture, whether or not the bare electron-phonon coupling is screened effectively by the free carriers depends on whether the evolution of the band induced by the phonon displacement pattern leads to a spatial variation of the density of charge carriers. In this regard, the situation can be very different depending on whether one or more valleys are populated.

To understand why, consider a phonon near $\Gamma$ ($\boq \to 0$) as a frozen atomic displacement with very long, but finite, wavelength. The static displacement associated to such a phonon shifts the valleys up and down in energy locally in space, following the local amplitude of the long-wavelength \Ag\ mode. In the adiabatic limit, the population of the electronic states will change to keep the system in the lowest energy state. Now, if only a single valley is occupied (say the K valley in the conduction band), the only possibility for the free carriers is to follow the spatially varying position of the bottom of the valley to keep the Fermi level constant (or more appropriately the electrochemical potential, since we are considering the atomic displacement as frozen). This implies a spatial variation of the charge of the free carriers that self-consistently screens the potential of the lattice displacement, thereby strongly suppressing electron-phonon coupling.

When at least two valleys are occupied, the situation can be very different if the atomic displacements associated with the phonon mode lead to a relative out-of-phase energy shift. This is indeed so in monolayer TMDs for the Q and K valleys in the conduction band and for the K and $\Gamma$ valley in the valence band (see Fig.~\ref{fig:band_ph}). In this case, in the presence of the atomic displacement (and when electrons/holes have the time to relax their energy and momentum during the phonon perturbation, i.e. in the adiabatic limit)  charge transfer between the valleys must also occur locally in real space for the system to stay in the ground state. Since part of the charge is now involved in this ``local'' process, the charge that can be displaced to screen the spatially varying potential generated by the phonon is reduced, hindering the effectiveness of screening. In other words, in the presence of two valleys that shift out-of-phase in response to long-wavelength phonon excitations, screening becomes much less effective in reducing the strength of the electron-phonon coupling.

This is the main mechanism responsible for the strong softening of the \Ag\ mode that we observe in the Raman measurements in all cases when two valleys are simultaneously occupied. For this mechanism to have a strong effect, it is essential that the phonon mode considered causes an out-of-phase shift in the the energy of the two valleys, because this maximizes the amount of inter-valley charge transfer that occurs locally in real space. For the \Eg\ mode detected in our Raman measurements on semiconducting TMDs, for instance, this is not the case: the long-range Fr\"ohlich potential associated with the \Eg\ phonon near $\Gamma$ shifts the energy of the two valleys in the same direction and by comparable amounts so that no (or minimal) inter-valley charge transfer occurs locally. As a result, despite the presence of multiple inequivalent valleys, the response of the system to \Eg\ phonons is similar to the canonical case with one single valley populated, with a strong electrostatic screening that suppresses the electron-phonon coupling.

It follows from these considerations that the underlying physical processes responsible for the enhancement of electron-phonon coupling in the presence of simultaneously populated inequivalent valleys is of quite general validity. As such, it is likely to occur in a multitude of materials and to have different experimental manifestations. It is quite remarkable that such a mechanism had not been identified earlier. We believe that this is because the analytic theoretical models commonly employed to treat electron-phonon interaction consider only one family of carriers (i.e., one single valley) coupled to phonons, and in these models the mechanism is simply not present.

Superconductivity provides an example of a physical phenomenon for which the enhanced electron-phonon coupling mechanism that we have identified may be at play. It appears that much of the existing data on gate-induced superconductivity observed in some of the TMDs  (e.g., \ws\ and \mos) is consistent with having the superconducting instability setting in only when multiple valleys are occupied. Having multiple valleys occupied would explain, for instance, why gate-induced superconductivity in \ws\ starts occurring at a lower electron density than in \mos~\cite{ye_superconducting_2012,lu_full_2018,costanzo_gate-induced_2016}, because, as we discussed here above, the energy difference between the K and Q valley is smaller in \ws\ than in \mos. It also explains why so far gate-induced superconductivity in semiconducting TMDs has been observed only upon electron accumulation and not with holes~\cite{ye_superconducting_2012,jo_electrostatically_2015}: this is because with the hole concentrations that can be reached by field-effect doping only the K or $\Gamma$ valleys are filled, so that the strength of the electron-phonon coupling remains weak. Indeed, in NbSe$_2$ --a metal having a similar valence band structure as WS$_2$ and MoS$_2$, but the Fermi level very deep relative to the top of the valence band~\cite{Lebegue2009,Nakata2018}-- robust superconductivity is observed~\cite{Ugeda2016,Xi2016,Tsen2016} and both the K and the $\Gamma$ valleys are occupied. The multi-valley mechanism responsible for the enhancement of the electron-phonon coupling is likely at play for other superconducting materials, with MgB$_2$ providing a notable example. Indeed, also in MgB$_2$ multiple inequivalent valleys are known to be simultaneously occupied and to move out-of-phase in energy when atoms are displaced according to a specific phonon pattern~\cite{An2001}. More work is  needed to fully substantiate the relevance of multivalley population for the occurrence of superconductivity in all these systems, but the known experimental phenomenology does provide significant circumstantial evidence supporting such a scenario.

\section{conclusions}

In conclusion,  we have  identified a strong and systematic electron-hole asymmetry in the vibrational properties of mono and bilayers of semiconducting TMDs, by performing Raman measurements on ionic-liquid-gated transistors. The experiments very systematically show that, in these systems, out-of-plane modes --such as the \Ag\ mode near zone center-- soften significantly as the concentration of free electrons in the conduction band is increased, while they are left unaffected upon hole accumulation. By performing a theoretical analysis based on first-principles simulations, we have revealed that the phonon softening originates from a previously unidentified mechanism that amplifies pronouncedly the strength of electron-phonon coupling when two inequivalent valleys are simultaneously populated. This mechanism explains the asymmetry in phonon softening observed experimentally in the Raman spectroscopy measurements as due to a much larger energy separation between valleys in the valence bands as compared to the conduction band. As a result, in the range of carrier densities accessed in the experiments, simultaneous occupation of multiple valleys occurs only upon electron accumulation and not for holes.

It is surprising that this mechanism had not been identified earlier. This is most likely due to the fact that analytic theoretical models used in the discussion of electron-phonon interactions commonly treat the case of a single family of carriers (i.e., a single valley) that is coupled the lattice vibrations. It is nevertheless clear from our discussion that the microscopic processes responsible for the enhancement of electron-phonon interaction in multi-valley systems are physically robust. Hence, the same mechanism should be at play in many different material systems (i.e., not only in semiconducting transition metal dichalcogenides), and manifests itself in a variety of physical phenomena. This is likely the case, for example, of superconductivity. Existing experiments on transition metal dichalcogenides appear to indicate that the occurrence of superconductivity --either gate-induced in MoS$_2$ and WS$_2$, or intrinsically present in NbSe$_2$--  correlates systematically with having multiple valley populated. Similar considerations can be made for other materials, with the case of MgB$_2$ being particularly notable. These examples underscore how the identification of a previously unappreciated mechanism enhancing the strength of electron-phonon interaction can have an important impact in the interpretation of many different phenomena observed experimentally.

\acknowledgements

We sincerely thank Francesco Mauri, Matteo Calandra, Haijing Zhang, and J\'er\'emie Teyssier for fruitful discussions. We gratefully acknowledge Alexandre Ferreira for continuous and precious technical support. A.F.M. gratefully acknowledges financial support from the Swis National Science Foundation (Division II) and from the EU Graphene Flagship project. M.G.\ and N.U.\ acknowledge support from the Swiss National Science Foundation through the Ambizione program. Simulation time was provided by CSCS on Piz Daint (project ID s825) and by PRACE on Marconi at CINECA, Italy (project ID 2016163963).


\renewcommand\theequation{\thesection\arabic{equation}}
\renewcommand\thefigure{\thesection\arabic{figure}}
\setcounter{equation}{0}
\setcounter{figure}{0}
\setcounter{table}{0}
\setcounter{section}{0}

\appendix

	\section{Methods}
    \label{app:methods}

Te realization of the field-effect transistors used to perform the experiments described in the main text relies on a combination of conventional nanofabrication processes and of techniques that are commonly employed to manipulate atomically thin crystals, \ie{} 2D materials. Monolayers and bilayers of MoS$_2$, WS$_2$ and WSe$_2$ are obtained by mechanical exfoliation of bulk crystals and transferred onto Si/SiO$_2$ substrates. The layer thickness is identified by looking at optical contrast under an optical microscope, and confirmed by photoluminescence measurements. To attach  electrical contacts to the atomically thin layers, a conventional process based on electron-beam lithography, electron-beam evaporation and lift-off is used (together with the electrodes, also a large side gate electrode is deposited). The contacts consist of Pt/Au evaporated films for WSe$_2$ and Au evaporated films for WS$_2$ and MoS$_2$, and are annealed at 200\degree{}C in a flow of Ar:H$_2$ gas (100:10 sccm). As a last step, the ionic liquid is drop-casted on top of the structure and confined under a 50 \mum{} thick glass plate to ensure a sharp optical image during the optical measurements. As rapidly as possible after depositing the liquid, the device is mounted in the vacuum chamber with optical access (\emph{Cryovac} KONTI cryostat)  where the measurements are performed (p $\approx$ 10$^{-6}$ mbar). The devices are left under vacuum at least overnight prior to the application of any gate bias to ensure that moisture or oxygen possibly present are pumped out of the system. The ionic liquid used in all experiments is 1-butyl-1-methylpyrrolidiniumtris(penta-fluoroethyl) trifluorophosphate [P14][FAP]. All measurements discussed here have been performed at room temperature; at lower temperature the same procedure cannot be applied because of freezing of the [P14][FAP] ionic liquid.

The Raman scattering spectroscopy measurements are performed using a home tailored confocal microscope in a back-scattering geometry, \ie{} collecting the emitted light with the same microscope used to couple the laser beam (at 514.5 nm) onto the device (see Fig.~\ref{fig:setup}c for a sketch of the experimental setup). The Back-scattered light is sent to a Czerni-Turner spectrometer equipped with a 1800 groves/mm grating which resolves the optical spectra with a precision of 0.5 cm$^{-1}$. The light is detected with the help of a N$_2$-cooled CCD-array (\emph{Princeton Instruments}). In the experiments, the gate voltage is slowly swept from positive to negative values. At V$_{\rm G}$ intervals of 0.2 V the sweep is interrupted and Raman spectra are recorded at fixed  V$_{\rm G}$ values. After reaching the largest negative gate voltage, V$_{\rm G}$ is swept back to positive values, and spectra are again recorded at fixed intervals to ensure that the experimental data are free from hysteresis. Finally, we have checked that the ionic liquid does not possess any characteristic Raman feature in the spectral range of interest.

DFT calculations are performed using Quantum ESPRESSO~\cite{Giannozzi2009,Giannozzi2017}, including a cutoff of the Coulomb interaction to implement the correct  boundary conditions and gates to simulate doping~\cite{Sohier2017}. We use a symmetric double-gate geometry;  for monolayers a single-gate geometry leads to minimal changes on the results. The exchange-correlation functional is approximated using the Perdew-Burke-Ernzerhof (PBE) form~\cite{PBE} of the generalized-gradient approximation.
To compute full phonon dispersions, we use the SSSP Accuracy library~\cite{prandini_standard_2018,Prandini2018} (version 0.7), a $60$ Ry energy cutoff, and a $32 \times 32 \times 1$ grid with $0.02$ Ry Marzari-Vanderbilt broadening~\cite{Marzari1999}. We then use density-functional perturbation theory~\cite{Baroni2001} --including Coulomb cutoff and gates-- to compute the phonons on a $12\times12\times1$ grid in the Brillouin zone.
For zone-center phonon calculations, we significantly reduce smearing and increase the sampling accordingly: Namely, we use a dense  $120 \times 120\times1$ k-point grid and Fermi-Dirac smearing corresponding to room temperature (0.002 Ry). Spin-orbit interaction is included by using fully relativistic norm-conserving pseudopotentials from the Pseudo-Dojo library~\cite{vanSetten2018}. Phonopy~\cite{Togo2015} is then used to generate the atomic-displacement patterns and to post-process forces to obtain phonon frequencies.

\section{Justification of Eq.~\eqref{eq:Dom} in the main text}
\label{app:softening}

In the the following, we show how to use expressions derived in Refs.~\cite{Lazzeri2006,Saitta2008,Calandra2010} to obtain Eq.~\eqref{eq:Dom} with the help of few assumptions valid in the case of semiconducting TMDs.
We are interested in the doping-induced frequency variations of a phonon mode at $\Gamma$ (phonon momentum $\boq \to 0 $) with respect to its value in the neutral case $\Delta \omega = \omega_{\rm{doped}}-\omega_{\rm{neutral}}$. The frequency of a given phonon mode corresponding to an eigenvector $\ket{\varepsilon}$ of the dynamical matrix at $\Gamma$, $\mathcal{D}(\boq=0)$, can be obtained as $\omega^2=\bra{\varepsilon}\mathcal{D}(\boq=0)\ket{\varepsilon}/ \langle\varepsilon|\varepsilon\rangle = \mathcal{D}_\Gamma/M$, where $M = \langle\varepsilon|\varepsilon\rangle$  is an effective mass associated with the phonon mode and $\mathcal{D}_\Gamma = \bra{\varepsilon}\mathcal{D}(\boq=0)\ket{\varepsilon} = \bra{u}\mathcal{C}(\boq=0)\ket{u}$ is the projection of the dynamical matrix on the corresponding eigenvector or equivalently of the force constants $\mathcal{C}$ on the physical displacements $\ket{u}$ of the phonon mode (with $\langle u|u\rangle=1$).
Assuming that the variation of the frequency is small compared to the  neutral frequency $\Delta \omega \ll \omega_{\rm neutral}$, we can  differentiate
($d\omega \approx d\omega^2/2\omega$) to obtain the following simple relation:
\begin{align}
\Delta \omega \approx \frac{\Delta \mathcal{D}_\Gamma}{2M\omega_{\rm neutral}}.
\end{align}
Neglecting the variations with doping of the contributions from the second-order derivatives of the ionic potential and the exchange-correlation kernel with respect to the atomic displacements in Eq.~(3) of Ref. \cite{Saitta2008}, we write:
\begin{align}
\Delta \mathcal{D}_\Gamma & \approx \Delta \mathcal{F}_\Gamma
\end{align}
where $\mathcal{F}_\Gamma$ is the part of the dynamical matrix at the $\Gamma$ point involving the electronic response to the phonon perturbation, and thus electron-phonon interactions.
The explicit expression for $\mathcal{F}_\Gamma$ depends on the adiabaticity of the system.
In the adiabatic limit relevant for TMDs, it reads~\cite{Saitta2008,Lazzeri2006,Calandra2010}:
\begin{align} \label{eq:FA}
\mathcal{F}_{\Gamma}&=\frac{1}{N_{\bok}}\sum_{\bok, n \neq m} |\langle\bok m|\Delta V_{\Gamma, \nu}|\bok  n\rangle |^2 \frac{f_{\bok, m}-f_{\bok, n}}{\varepsilon_{\bok, m}-\varepsilon_{\bok, n}} \\
& +\frac{1}{N_{\bok}}\sum_{\bok, n} |\langle\bok n|\Delta V_{\Gamma, \nu}|\bok n\rangle |^2 \frac{\partial f}{\partial\varepsilon}\Big|_{\varepsilon_{\bok,n}}
\end{align}
where $\Delta V_{\boq, \nu}$  is the perturbation of the effective potential due to phonon $(\boq, \nu)$, $f_{\bok, n}$ is the occupation of the state $|\bok n\rangle$ corresponding to the band $n$ at crystal momentum $\bok$ with energy $\varepsilon_{\bok, n}$.
In the first term, there are only interband contributions ($n \neq m$) between occupied and empty states. We have verified that the numerator is much smaller than the denominator $|\varepsilon_{\bok, m}-\varepsilon_{\bok, n}|$, which is equal to the direct gap in the neutral case, so that this term can be neglected. The second term involves intraband transitions between states close to the Fermi surface and it is non-zero only in doped systems, while vanishing in neutral TMDs. This is the main contribution to doping-induced variations, so that we can write:
\begin{align}
\Delta\omega &\approx \frac{\Delta \mathcal{F}_{\Gamma}}{2M \omega_{\rm neutral}}  
\approx \frac{1}{N_{\bok}}\sum_{\bok, n} \frac{|\langle\bok n|\Delta V_{\Gamma, \nu}|\bok n \rangle |^2}{2M \omega_{\rm neutral}} \frac{\partial f}{\partial\varepsilon}\Big|_{\varepsilon_{\bok,n}}\notag\\
&=  \frac{1}{N_{\bok}}\sum_{\bok, n} |g^{(n)}_{\nu,\bok,\bok}|^2 \frac{\partial f}{\partial\varepsilon}\Big|_{\varepsilon_{\bok,n}},
\end{align}
where we have introduced the square electron-phonon matrix element due to intraband transitions:
\begin{equation}
|g^{(n)}_{\nu,\bok,\bok}|^2  = \frac{|\langle\bok n|\Delta V_{\Gamma, \nu}|\bok n\rangle |^2}{2M \omega_{\rm neutral}}~.
\end{equation}
The derivative of the Fermi distribution is negative (the phonon frequency softens) and non-zero only in a small energy window of order $k_B T$ around the chemical potential, so that only electronic states belonging to the temperature-smeared Fermi surface contribute. In the zero temperature limit the derivative becomes a delta function at the Fermi energy $\varepsilon_F$ and we obtain
\begin{align}\label{eq:deltaOmF}
\Delta\omega &\approx - \frac{1}{N_{\bok}}\sum_{\bok, n} |g^{(n)}_{\nu,\bok,\bok}|^2 \delta(\varepsilon_{\bok,n}-\varepsilon_F) \notag\\
&= - N(\varepsilon_F) \langle g^2_{\nu}\rangle_{FS},
\end{align}
which is Eq.~\eqref{eq:Dom} of the main text. Here we have introduced the density of states at the Fermi energy
\begin{equation}
N(\varepsilon_F) = \frac{1}{N_{\bok}}\sum_{\bok, n}  \delta(\varepsilon_{\bok,n}-\varepsilon_F)~,
\end{equation}
and the square electron-phonon coupling averaged over the Fermi surface:
\begin{equation}\label{eq:g2aver}
\langle g^2_{\nu}\rangle_{FS} = \frac{\sum_{\bok, n} |g^{(n)}_{\nu,\bok,\bok}|^2 \delta(\varepsilon_{\bok,n}-\varepsilon_F)}{\sum_{\bok, n}  \delta(\varepsilon_{\bok,n}-\varepsilon_F)}~.
\end{equation}


\begin{thebibliography}{10}
\bibitem{capa}%
  \BibitemOpen
  \bibinfo {note} {The values of carrier density are estimated using values of
  the capacitance per unit area obtained from previous experiments (in which
  the capacitance was obtained using the Hall effect to measure the density of
  accumulated charge). These past experiments show that in ionic-liquid gated
  transistors on mono and bilayer semiconducting TMDS --in the range of density
  explored here-- the capacitance value is largely determined by the density of
  states (i.e., taking into account the quantum capacitance is essential).
  Since the density of states is different for mono and bilayers and --once the
  thickness is fixed-- for the conduction and valence bands, different values
  for $C$ have to be used. At room temperature, the capacitance does not depend
  much on the specific compound considered and here we use the values measured
  on WSe$_2$ mono and bilayers, which are: C$_{\rm ML}^e$ = 7.3 \uFsqcm ;
  C$_{\rm ML}^h$ = 17 \uFsqcm ; C$_{\rm BL}^e$ = 22.8 \uFsqcm ; C$_{\rm BL}^h$
  = 24 \uFsqcm \ (the subscript ML and BL refer to mono and bilayer; the
  superscript $e$ and $h$ indicates the value to use for accumulation of
  electrons and holes, respectively).}~%
  \bibAnnoteFile{Stop}{capa}%
\bibitem{2LAm}%
  \BibitemOpen
  \bibinfo {note} {In view of the microscopic complexity of the resonant
  process from which the 2LA(M) Raman peak originates, and of the fact that
  --as it will be discussed explicitly in Section~\ref {sec:theory}-- our main
  interest here is on long-wavelength phonons, we will not discuss the behavior
  of the 2LA(M) Raman peak. As a only remark, note that --just like for the \Ag
  \ mode-- the LA(M) mode corresponds to an out-of-plane motion of the atoms in
  the lattice, and a finite electron-phonon coupling is present, as
  demonstrated by the sensitivity of the mode frequency to the introduction of
  charge carriers.}~%
  \bibAnnoteFile{Stop}{2LAm}%
\end{thebibliography}
\end{document}